# Quantitative Susceptibility Mapping in Cognitive Decline: A Review of Technical Aspects and Applications

Shradha Verma[1], Tripti Goel[1]*, and M Tanveer[2]

[1]*Biomedical Imaging Lab, National Institute of Technology Silchar, 788010, Assam, India.

[2] Discipline of Mathematics, Indian Institute of Technology Indore, Simrol, 453552, Madhya Pradesh, India.

*Corresponding author(s). E-mail(s): triptigoel@ece.nits.ac.in;  Contributing authors: shradha-rs@ece.nits.ac.in ; mtanveer@iiti.ac.in;

*Abstract*

**Background:** In the human brain, essential iron molecules for proper neurological functioning exist in transferrin ($tf$) and ferritin ($Fe3+$) forms. However, its unusual increment manifests iron overload, which reacts with hydrogen peroxide. This reaction will generate hydroxyl radicals, and iron's higher oxidation states. Further, this reaction causes tissue damage or cognitive decline in the brain and also leads to neurodegenerative diseases. The susceptibility difference due to iron overload within the volume of interest (VOI) responsible for field perturbation of MRI and can benefit in estimating the neural disorder.

**Method:** The quantitative susceptibility mapping (QSM) technique can estimate susceptibility alteration and assist in quantifying the local tissue susceptibility differences. It has attracted many researchers and clinicians to diagnose and detect neural disorders such as Parkinson's, Alzheimer's, Multiple Sclerosis, and aging. The paper presents a systematic review illustrating QSM fundamentals and its processing steps, including phase unwrapping, background field removal, and susceptibility inversion.

**Results:** Using QSM, the present work delivers novel predictive biomarkers for various neural disorders. It can strengthen new researchers' fundamental knowledge and provides insight into its applicability for cognitive decline disclosure.

**Conclusion:** The paper discusses the future scope of QSM processing stages and their applications in identifying new biomarkers for neural disorders.

*1. Introduction*

Quantitative susceptibility mapping (QSM) is a post-processing neuroimaging technique [1, 2] employed to determine the susceptibility difference within the local tissue. It offers the 3D mapping of the susceptibility of brain tissues which often practices in identifying biomarkers for various neurological diseases.

Iron plays a vital role and is distributed heterogeneously in the brain region. The presence of iron is dynamic as its existence increases with age and decreases when the diet lacks iron content [3, 4]. Iron is responsible for various neurological functions [5, 6], such as myelin building, proper neurotransmitter functioning, oxygen movement, protein synthesis, and cell growth. The two essential iron molecules in the brain are transferrin iron transporter that carries iron from blood to brain tissues via receptors. In contrast, another iron molecule, ferritin is responsible for excessive iron storage (that is not instantly involved in metabolic activities).

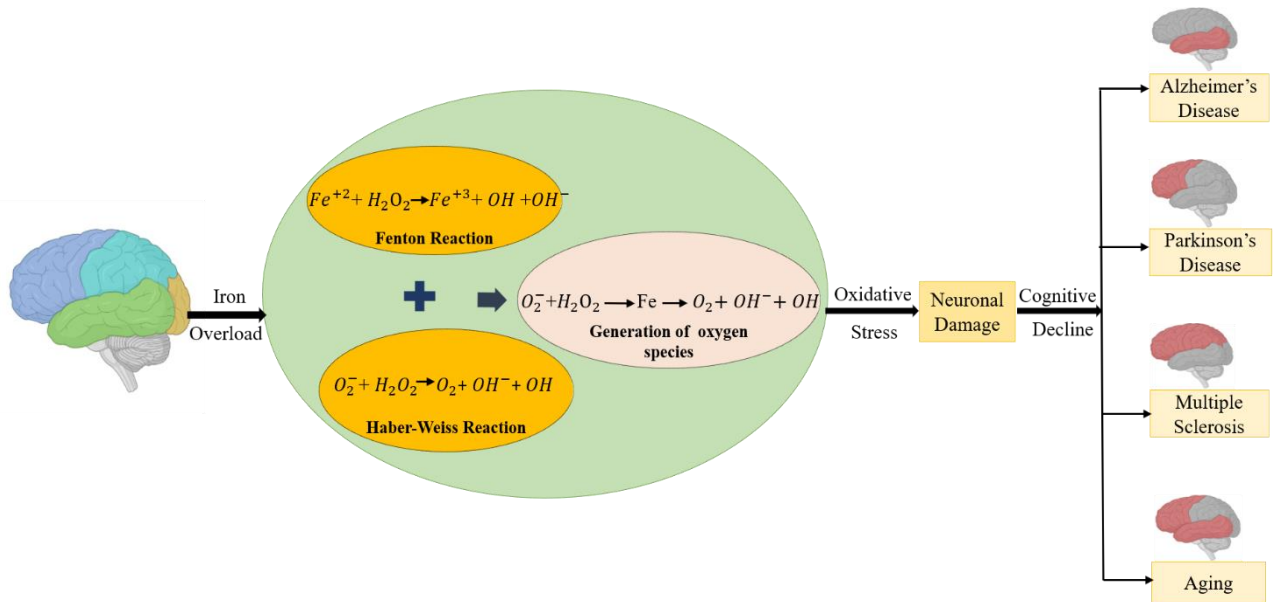

Fig.1 Cognitive decline in the brain due to iron overload

When this iron reacts with hydrogen peroxide, it generates hydroxyl radicals, and its higher oxidation state is termed as the Fenton reaction. Similarly, hydroxyl radicals with oxygen species are obtained in the Haber-Weiss reaction when oxygen atoms react with hydrogen peroxide. These two reactions are responsible for the reactive oxygen species that affect the different brain regions, resulting in the brain's malfunctioning, as shown in Fig.1.

Moreover, the ferritin content in the brain alters with the magnetic field variation [7]. Particularly, indirect iron quantity is affected by the magnetic field. Ferritin is the protein that stores this iron and delivers it according to the human body's requirements. Also, other biometals, like calcium, magnesium, etc., influence the cellular structure of the human brain. According to Haacke et al. [8], primary iron deposition in the deep gray matter (GM) region can be determined by phase contrast. However, iron presence is also observed in the brain's other parts, and research is still going in that direction. Excessive iron is affected by the total magnetic field. It induces frequency shifts in magnetic resonance (MR) signals because Larmor frequency depends on the echo time, gyromagnetic ratio, coil sensitivity, and magnetic field variation. Brain tissues show diamagnetic or paramagnetic properties. Therefore, gets disturbed in the presence of field variation and causes an alteration in overall magnetization. When this magnetization is convoluted with a point-wise dipole function in the fourier domain, it yields magnetic field variation, which is proportional to susceptibility value (mathematical equation mentioned in the foundation of QSM).

Patients with neurodegenerative diseases usually contain a higher iron level than normal controls. Different brain regions show a different level of iron content, including the White Matter (WM) and GM, Putamen (PUT), Cerebrospinal Fluid (CSF), Caudate Nucleus (CN), Basal Ganglia (BG), Substantia Nigra (SN) and others. An imbalance in iron level perturbates the homeostasis of the healthy brain and causes cognitive impairment. Further, that is responsible for neural disorders like Alzheimer's disease (AD), Parkinson's disease (PD), and others [9, 10]. Neuroimaging techniques can be employed for detection of these neural disorders. The multi-gradient echo (GRE) sequence of magnetic resonance imaging can investigate the local susceptibility difference in different brain regions. Moreover, the GRE sequence of QSM provides complex data. However, phase data prefers over magnitude because phase data contain more information than magnitude. Still, direct phase mapping fails to provide the desired image; therefore, the QSM technique finds application to diagnose biomarkers for early diagnosis of neurodegenerative disease. Because of the field variation of susceptibility maps, QSM can detect and diagnose neurological disorders. Also, recent literature has demonstrated that QSM images are advantageous in handling a variety of neuroimaging difficulties.

This paper is organized as follows; section 2 addresses the search strategy. Section 3 introduces the principle of QSM, which consists of the foundation of QSM, data acquisition, and processing techniques like the phase unwrapping method, background field removal, and susceptibility inversion. Furthermore, the application of QSM is included in section 4. It indicates the technique's effectiveness as a biomarker for neural disorders. Lastly, in section 5 and section 6, the future scope and the conclusion are added, respectively.

## 2. Search Strategy

We searched the papers from PubMed https://pubmed.ncbi.nlm.nih. gov/ and Google Scholar https://scholar.google.com/ using keywords "QSM", "Cognitive decline" and "Applications of QSM", which searched a total of 550 articles. In the first screening, 350 writings are rejected based on abstracts and topics which did not match the current modality and clinical applications. The second screening has been done and eliminated 112 previous works due to the distinct modalities and other types of applications. Finally, 88 papers are added in this paper, which addresses the QSM pipeline perception and encourages researchers to identify the biomarker for cognitive decline. The flow chart of search strategy is shown in Fig. 2.

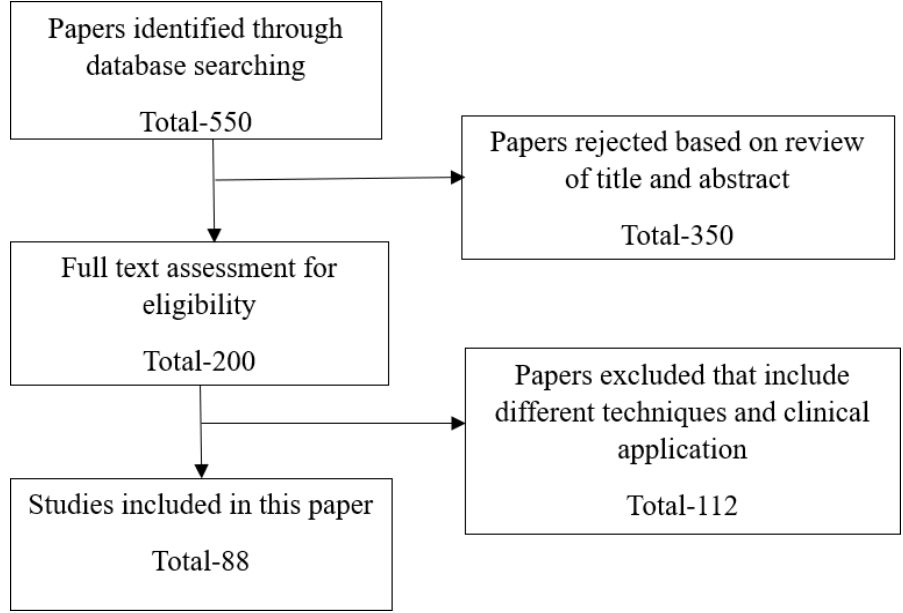

Fig. 2 Flowchart of search strategy

*3. Foundation of QSM*

The biological tissues of the human brain are diamagnetic in nature [11]. Other metals like calcium, iron, fluorine etc., however, exhibit paramagnetic properties whose susceptibility lies in the range of $\pm 10^{-5}$ to $\pm 10^{-6}$. Moreover, the energy levels of molecules depend on the electron's spin and angular momentum. According to Boltzmann's distribution [12], an effective magnetic moment will be altered with the number of an unpaired electron and leads to a change in magnetization. In addition, this magnetization is equivalent to measuring the alteration in magnetic susceptibility [13], a second-order tensor term that has a dependency on the magnetic field and magnetic hysteresis.

Magnetic resonance imaging (MRI) is a non-invasive technique that yields multi-dimensional anatomy images using magnetic properties. When the human brain is exposed to the static magnetic field of MRI, disturbance appears in atoms or molecules of the brain due to the applied magnetic field which is termed as the demagnetization field.

Total magnetic field $B(\vec{r})$ with arbitrary point $\vec{r}$ corresponds to summation of static field $B_{ext}(\vec{r})$ and demagnetization field $B_{demag}(\vec{r})$. Complete demagnetization field can be defined in the term of the Lorentz distant field $B_{dist}(\vec{r})$ and near field $B_{near}(\vec{r})$. Mathematically, first order Lorentz distant field for non-ferromagnetic material is formulated as:

$$B_{dist}(\vec{r}) = B_{ext}(\vec{r}) \cdot \int_{V_{dist(\vec{r})}} \chi_{apparent}(\vec{r'}) \cdot b_\chi(\vec{r} - \vec{r'}) d^3\vec{r'} \qquad (1)$$

Here, $b_\chi(\vec{r}) = \dfrac{3\hat{r}(\hat{z}.\hat{r}) - \hat{z}}{4\pi . \|\vec{r}\|_2^3} . \hat{z} = \dfrac{3\cos^2\theta - 1}{4\pi . \|\vec{r}\|_2^3}, \vec{r} \neq 0$

where, $b_\chi(\vec{r})$: unit dipole function, $\chi_{apparent}$: apparent magnetic susceptibility at point $\vec{r}$, $V_{dist}(\vec{r})$: distant region from hydrogen nucleus at $\vec{r}$ location, $b_\chi(\vec{r} - \vec{r'})$: shifted dipole function from $\vec{r}$ to $\vec{r'}$ location, $\hat{r}$: unit vector in the direction $\vec{r}$, $\hat{z}$: unit vector in the z-direction, $\|.\|_2$: euclidean norm and dot indicate point-wise product.

As the $B_{dist}(\vec{r})$ associated with macroscopic magnetization, therefore, it can define in magnetization term i.e., $M(\vec{r})$. Function like dipole function $b_\chi(\vec{r})$ expression is same as green's function $G(\vec{k})$ which can be directly employed to solve the magnetic field equations. In addition, on applying the condition $\chi \ll 1$, product of magnetic permeability and magnetization ($\mu_o M(\vec{r})$) can be replaced with the multiplication of magnetic field and susceptibility (i.e., $\mu_o M(\vec{r}) \approx B_o \chi(\vec{r})$). Thus, $B(\vec{r})$ in frequency domain can be defined as:

$$B(\vec{r}) = B_o \, FT^{-1}(\chi(\vec{k}) \cdot G(\vec{k})) \tag{2}$$

where $FT^{-1}$: inverse fourier transform, $\chi(\vec{k})$: fourier domain susceptibility, $\vec{k}$: coordinate vector in fourier space, $B_o$: static magnetic field. MR signal can be expressed with Larmor frequency $f_L(\vec{r})$ related to field map and it is written in equation form as:

$$f_L(\vec{r}) = -\frac{\gamma}{2\pi} \cdot B(\vec{r}) \tag{3}$$

where $\gamma$: gyromagnetic ratio. At the demodulation frequency $f_R$, transformation can occur from laboratory $f_L$ to rotating frame $\Delta f_{MR}$ and forced Equation (3) to depend on parameters likes phase offset and chemical shift frequency. When the external magnetic field in controlled by homogeneous ($B_{homo}$) and inhomogeneous ($B_{nhomo}$) magnetic field. Chemical shift and phase offset (i.e, ($f_o - f_R$)) constant value could be considered. In the equation form, can be formulated as:

$$\Delta f_{MR} \approx -\frac{\gamma}{2\pi}(B_{nhomo}(\vec{r}) + B_o(\chi_{apparent} \otimes d_\chi)) \tag{4}$$

where, $d_\chi$: modified microscopic unit dipole function.

Susceptibility measures the degree of magnetization in the presence of the applied magnetic field as changes in susceptibility are due to the MR signal. The GRE sequence [11] investigated local magnetic susceptibility, which records the frequency shifts. $T2^*$ weighted sequence is required for susceptibility-weighted imaging (SWI) [14,15]. Specifically, phase and magnitude image information are needed to obtain an SWI image, but it failed to quantify the susceptibility difference. Therefore, QSM images are needed to avoid the demerits of SWI. QSM utilized pulse sequences as these sequences are sensitive to tissue susceptibility differences. In addition, the QSM is a post-processing technique that assists in quantifying the susceptibility of local tissue by observing the susceptibility variation in the brain.

Furthermore, quadrature detection generates magnitude and phase images [16] separately. Phase exhibits more details that are further post-processed. Final QSM image reconstruction from MRI data required many steps. The first step includes phase unwrapping, where phase discontinuity is observed whenever the phase value exceeds the $2\pi$ range. Also, it is responsible for aliasing and error effects in the final image. Therefore, phase unwrapping techniques are needed to mitigate the aliasing effects. Next, the background field removal technique can be performed as a second step. Magnetic susceptibility of brain regions other than the region of interest (ROI) creates artifacts. Hence, high-pass filtering or some other background removal techniques [17] are employed to mitigate the effect of the nonlocal field. Finally, susceptibility inversion is conducted, which involves a set of mathematical equations to reconstruct the QSM image.

Several studies have been done on QSM to identify susceptibility in neurodegenerative disease cases. Schweser et al. [18] have provided an overview of pre-processing steps for MRI phase data, including pre-processing steps assumptions and constraints. The paper summarized QSM solution strategies, investigated the reconstruction of magnetic field maps from multi-channel phase images, and discussed the background field correction approach. Similarly, Deistung et al. [19] have briefly summarized the theoretical foundations of QSM and susceptibility tensor imaging (STI). These approaches are employed to quantify the magnetic susceptibility of MRI phase data. Both techniques mitigated the ill-posed inverse problem of the QSM map by determining the magnetic susceptibility from local magnetic fields. In addition, the author's work introduced QSM methodology and clinical application. Wang et al. [20] have discussed the clinical importance of QSM in detecting higher susceptibility of biometals and summarised the neural disorders for which QSM can be utilized. Moreover, they investigated the robust and automated QSM model using the Bayesian approach. Finally, the author concluded using GRE sequences, automatic preservation of phase and magnitude data is possible. Hence, QSM can be employed as a routine MRI for imaging objectives. None of these studies systematically discussed the QSM processing pipeline steps and applications of neural disorders. In contrast, the current review paper describes the recent methodologies of QSM, including phase unwrapping, background field removal, and susceptibility inversion. Moreover, this paper also presented the systematic review and analysis of application of QSM to find the new biomarkers and region of interest (ROI) for diagnosis of various neural disorders. Besides, the paper elaborated on various challenges and the future scope. Table 1 shows the summary of the QSM review papers which are published so far.

Table 1 Comparison of published papers and ours review paper that address QSM overview

| References | Years | Included papers | Contribution |
| --- | --- | --- | --- |
| [18] | 1982-2015 | ~176 | Fundamental of QSM |
| [19] | 1984-2016 | ~265 | Basics of QSM and STI with clinical application |
| [20] | 1987-2017 | ~193 | Detect the susceptibility of bio-metal using QSM |
| Ours | 1990-2022 | 88 | Foundation of QSM, QSM reconstruction process, clinical application in detecting the biomarker and future scope. |

QSM processing steps are mentioned in the block diagrams, as shown in Fig.3. The upcoming subsection briefly discusses individual steps and different methodologies.

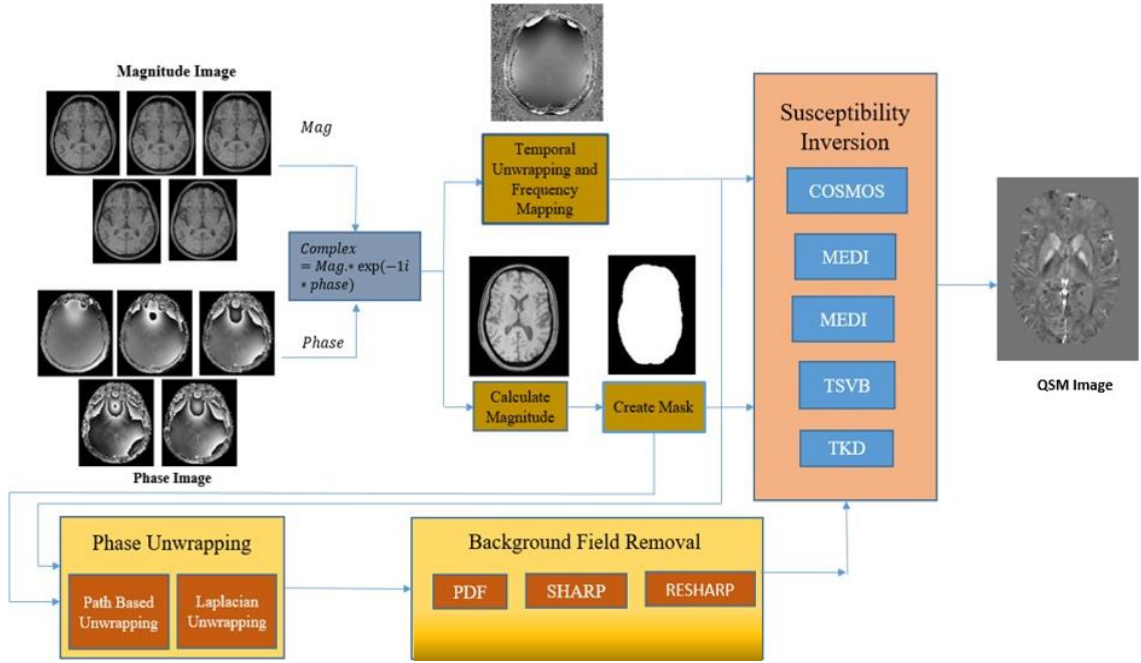

Fig. 3 Block diagram of QSM reconstruction process

*3.1 Phase Unwrapping*

The raw phase value has phase discontinuity or phase jumps whenever the phase limit goes beyond the $2\pi$ range. Therefore, an unwrapping technique is required to mitigate these discontinuity or errors. Fig. 4 (a) and (b) show the magnitude image and wrap phase image, respectively. Phase unwrapping algorithms [21] are categorized into path-following [22] and Laplacian-based unwrapping [23] as shown in Fig. 4(c) and (d), respectively.

The path-following method (PFM), unwrap the phase value whenever the phase difference between two samples exceeds the $2\pi$ threshold along the integral path. In this algorithm, $+2\pi$ and $-2\pi$ phase value is added or subtracted depends on positive and negative of phase jumps. And, this step is continued until phase errors is minimized or the correct phase values restored.

Laplacian unwrapping technique employs a second order differential operator to unwrap phase image. Technique removes discontinuities or phase jumps while maintaining smooth phase values. For the uniform region, the laplacian operator provide zero value, whereas non-zero phase value is obtained in the non-uniform region case. Laplacian unwrapping methodology, however, can alter both the background value as well as other values.

Ryu et al. [24] have proposed a deep learning (DL) concept for phase unwrapping. They have created a unique network using bidirectional recurrent neural network (RNN) modules. The proposed network learned global features and constructed a loss function based on the error of phase image contrast. The network delivered a satisfactory performance for simulated and in-vivo data. The network output demonstrated its ability to manage global phase wraps and generate unwrapped phase images. Similarly, Zhou et al. [25] have designed a PHU-NET to unwrap the phase data. The author included several pre-processing techniques to improve the model's robustness and efficiency. Broadly, they evaluated the approach quantitatively and found that for simulated data, it performed well even in low signal-to-noise conditions.

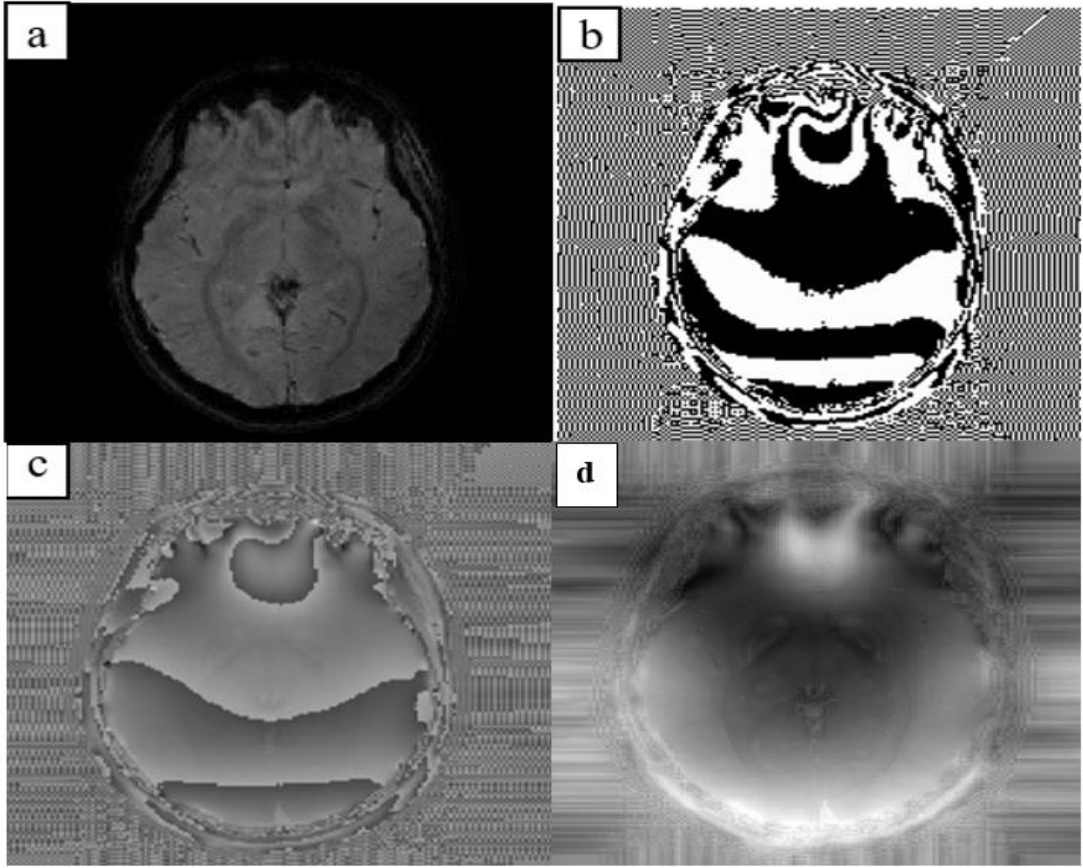

Fig. 4 Phase unwrapping Methods: (a) Magnitude image, (b) Wrapped phase image, (c) Path following method, (d) Laplacian method

*3.2 Background field Removal*

An image region of magnetic susceptibility that does not come under ROI or voxel of interest is considered a background field (shown in Fig. 5(a) and Fig. 5(b)). Background artifacts like air tissue etc., affect the local susceptibility value, resulting in poor phase mapping. Therefore, the removal of the background field is essential. The traditional high pass filter (HPF) method is not compatible as it is inefficient in eliminating the background field without affecting the local tissue susceptibility, as shown in Fig. 5. Some popular methods for background field removal include sophisticated harmonic artifact reduction for phase (SHARP) [26, 27], regularization-enabled SHARP (RESHARP) [28], and projection on dipole field (PDF) [29].

Schweser et al. [26] have proposed the SHARP technique for background field removal. SHARP is a filtering technique based on harmonic region (i.e., region which exists outside of ROI). In contrast, the non-harmonic internal region can be computed using the mean value property, as shown in Fig. 5(c). The technique provides satisfactory results; however, it yielded an underdetermined solution.

Local and background information is required to get the desired image solution. Hence, The least norm can be employed that assists in acquiring the local field. In contrast, the least-norm solution had a minimization issue. So, RESHARP [28] technique, which applies Tikhonov regularization, comes into the picture. It mitigates background noise, resulting in fewer artifacts and better tissue contrast, as shown in Fig. 5(d).

It is observed that the inner product of a background dipole's field outside the ROI and a local dipole's field inside the ROI is almost zero, except for local dipoles near the boundary. This insight is the foundation for the PDF method used to differentiate between the local and background fields. The methodology explains that the unit dipole of the background can calculate the total field. Further, this can be justified with the orthogonality theorem, which states that the inner product of each local unit dipole with any possible background unit dipole is zero, as shown in Fig 5(e).

Besides the conventional background field removal approach DL-based background field removal, approaches are discussed by Bollmann et al. [30]. In [30], the authors have demonstrated SHARQnet whose performance was better compared to traditional approaches. Without needing a brain masking step, this network can perform background removal. Moreover, the performance of SHARQnet was excellent as it acted close to the brain's boundary, yielding a substantially smaller error. Similarly, Liu et al. [31] suggested a background field removal using the padding technique. In the neural network's method, they used the neighboring valid voxels to calculate the invalid voxels of feature maps at volume boundaries. Studies employing simulated and in-vivo data demonstrate that padding significantly increases estimation accuracy and decreases artifacts in the results, which is advantageous in removing background fields and reconstructing single-step QSM.

Furthermore, the final step of the QSM pipeline, i.e., susceptibility inversion, is discussed in the upcoming subsection.

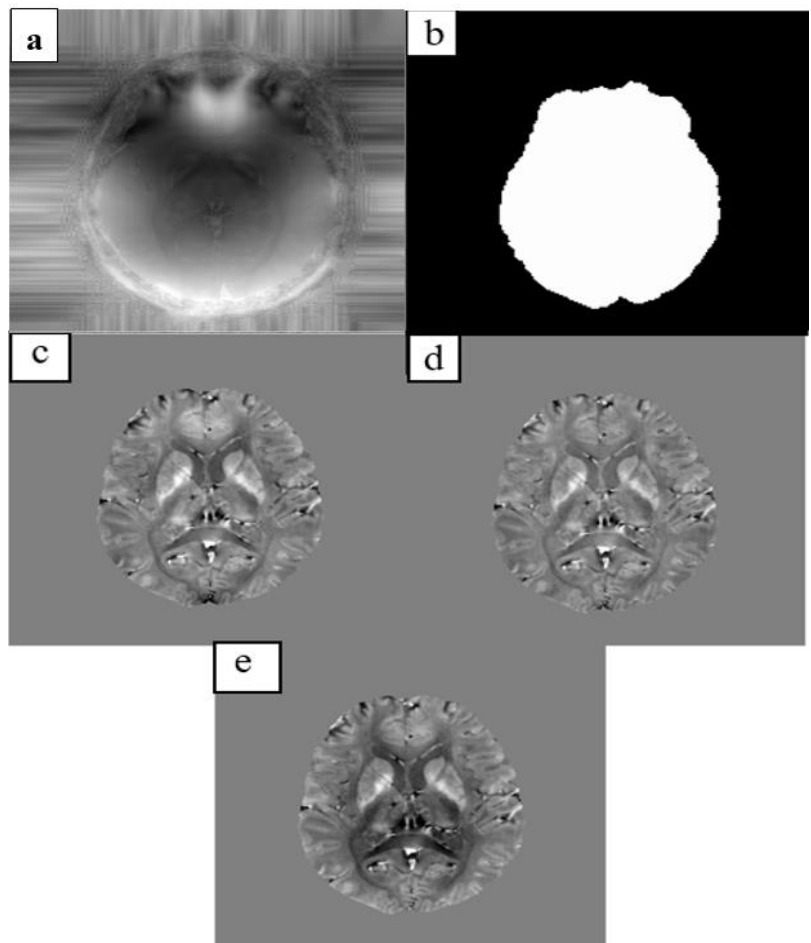

Fig. 5 Background Removal techniques:(a) Unwrapped phase image, (b) Mask, (c) SHARP, (d) RESHARP, (e) PDF.

*3.3 Susceptibility Inversion*

Variation of susceptibility at a magic angle leads to an ill-posed issue. To overcome this problem, susceptibility inversion is employed, whose objective is to acquire a desirable result using local field maps. Besides that, some inversion techniques are mentioned below:

Calculation of susceptibility through multiple orientation sampling (COSMOS) instead of taking a single orientation; this method is based on multiple orientations. The conic surface has two zeros at the magic angle (54.7º), and the zeros field map inversion does not provide the correct information in the frequency domain. To minimize this issue, the sampling operation can opt for discretized data that ignore the zeros [32]. Additionally, discrete data also led to ill-conditioned problems and amplified the noise level. To resolve this inversion problem, [33] article suggested the object's rotation with respect to the main magnetic field followed by resampling. And it should perform in such a manner that the Fourier domain coordinate axis remains attached to the object's frame of reference. However, two-orientation sampling is insufficient; therefore, third-angle orientation is required. These orientations help to take only the non-zero elements in the frequency domain, shown in Fig. 6(a).

In morphology-enabled dipole inversion (MEDI), the susceptibility map and magnitude image exploit the same GRE sequence for reconstructions. Because the voxel information of the susceptibility map is associated with magnitude image information [34, 35], it provides the benefit of minimizing the artifacts at the boundary. Mostly L1 norm minimization opts for this objective, whose result is shown in Fig. 6(b).

Total variation using split Bregman (TVSB) utilizes the L1 and L2 norms methods for reconstruction, as shown in Fig. 6(c). Regularized parameters λ and β of algorithms reduce the ringing effect and maintain the map's image contrast [36, 37]. However, over-regularization in TVSB output is still challenging to get a clear visible susceptibility difference map. A weighted total variation using split Bregman (WTVSB) can be employed for these, which weighs the data effectively and mitigates the noise effect. In addition, WTVSB also appears helpful in fast QSM reconstruction and in solving the 3D deconvolution problem.

Truncated singular value decomposition (TSVD) is an alternative method to least squares (LS). Strategies minimize the square mean value more effectively than LS algorithms, shown in Fig. 6(d). The closed-form solution of TSVD at k-value with few conditions provided the same outcome as the Tikhonov-regularized minimal norm [38, 39] method.

In the threshold k-space Division (TKD) [40], when an external field is applied, the overall magnetic field gets disturbed due to the susceptibility difference of a local field in the fourier domain, as discussed in the foundation part of QSM. Thus, it can be expressed as a susceptibility distribution that causes due to field perturbation. This methodology introduces a threshold parameter for the susceptibility inversion that truncates small or tends towards zero values in the k-space region. Primarily, the user defines a threshold value operated that successfully truncates small values in the k-space, as shown in Fig. 6(e).

Alternatively, the DL-based inversion technique can be opted for QSM reconstruction. Chen et al. [41] have implemented the QSMGAN technique, which uses the 3DUnet deep convolutional neural network. The method delivers lower edge discontinuity artifacts and better accuracy using patch-based neural networks and cropping techniques. Further, the image quality, training process stability, and accuracy improved by adding a wasserstein generative adversarial network+ the gradient penalty network. The paper result presented that the method for single orientation may show the same result as COSMOS. Its performance was found to be good. Similarly, in [42], the author has constructed QSMnet architecture for reconstruction. Basically, the network exhibited a modified U-net structure trained by using a COSMOS image. The performance of QSMnet is found out to be outstanding methodology in multi-head orientation cases.

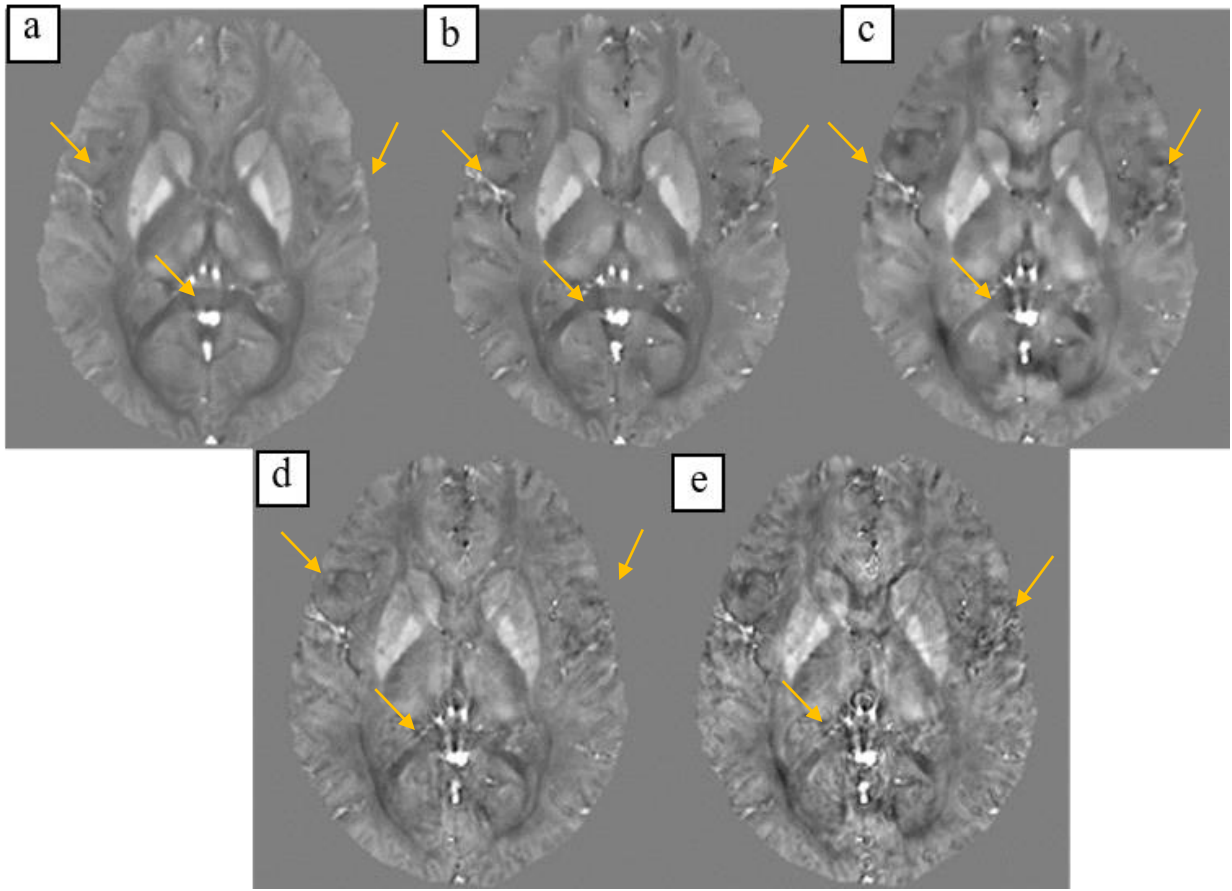

Fig. 6 Susceptibility Inversion (a) COSMOS (b) MEDI (c) TVSB (d) TSVD (e) TKD

The current study covers various techniques for QSM reconstruction that assist in understanding the mentioned topic. In the QSM pipeline, individual stages are discussed, encouraging the researcher to have insight into the QSM basics and their working. Further, these methodologies promote to detect the neurodegenerative disease. Therefore, the application of QSM is stated in the subsequent section for biomarker identification.

*4. Application of QSM*

*4.1 Alzheimer's Disease*

Alzheimer's Disease (AD) is caused due to an imbalance in the amount of protein that is developed inside and around the brain cell [9]. As introduced, iron overload causes oxidative damage that leads to neuronal disorder. Previous studies [43, 44, 45] also presented that excessive iron is responsible for depositing beta-amyloid and neurofibrillary tangles, which can cause AD. Because of that, metabolic complications are found in the cortex and the hippocampus (HP). A literature on brain regions for the higher iron accumulation in AD patients are illustrated in Table 1.

Table 1 Application of QSM in AD

| Citation | Method | Database | Resulted higher Iron accumulation regions |
|---|---|---|---|
| [46] | PDF and TKD | AD-8 HC-11 | PUT, Posterior GM and WM |
| [47] | MEDI and PDF | AD-27 HC-18 | PUT and CN |
| [48] | Variable SHARP and TKD | rTg4510 mice-10 Controls-10 | Corpus callosum, Striatum, HP and Th |
| [49] | MEDI and PDF | AD -57 | Precuneus and allocortex |
| [50] | - | AD -30 HC-30 | CN and PUT |
| [51] | - | AD-69 HC-25 | GP, PUT and BG |
| [52] | Variable SHARP and STAR-QSM | AD-23 HC-30 | Hippocampus fimbria |
| [53] | SHARP | AD-59 HC-22 | Cerebral veins, Th and DN veins |
| [54] | Least-squares estimation and Linear SVM | AD-61 HC-36 | Parietal lobe and GM |
| [55] | MEDI and PDF | AD-68 HC-43 | CN, PUT, GP and Th |

*Pulvinar Nucleus of the Thalamus*

Acosta et al. [46] have specified the magnetic susceptibility difference in deep brain nuclei, particularly in the brain regions of PUT, posterior GM, and WM regions. The study examined the hypothesis that there is an ideal parameter range for regularisation. Further, the paper discussed the $l1$-norm regularization parameter. Results showed relatively unattenuated quantitative values for AD patients and healthy controls (HC). However, the method performed on a small number of datasets; therefore, results may vary in other cases. Moon et al. [47] have determined the iron deposition pattern of vascular dementia (VaD) and AD in the brain's four regions: globus pallidus, PUT, CN, and pulvinar nucleus of the thalamus. The author used variance and post hoc analyses to compare VaD, AD, and normal control iron levels. Moreover, a partial correlation analysis was employed to evaluate the connections between the susceptibility value of age and cognitive decline. Results showed that susceptibility value difference is found to be higher in the PUT and CN regions of the VaD and AD individuals. However, the susceptibility values for VaD and AD are identical. In [48], a mouse model for in-vivo and ex-vivo has been examined that measures iron and myelin content in different brain regions. Here, magnetic susceptibility differences were identified in the corpus callosum, HP, and Th areas. Paper suggested that QSM could be employed for early-stage detection of improper tau balance.

In [49], the author has estimated the susceptibility changes of iron deposition and beta-amyloid using the voxel and ROI-based QSM. This susceptibility alteration was examined in 19 elderly cognitive normal (CN), 19 individuals with Mild cognitive impairment (MCI), and 19 AD subjects. Also, AD spectrum outcomes compared with grey matter volume (GMV) for neuronal loss. Furthermore, the finding of the technique suggested that QSM worked well in determining the difference among CN, MCI, and AD individuals compared to GMV changes. Similarly, Du et al. [50] have presented the susceptibility associated with the brain's iron. They demonstrated the correlation between susceptibility value and cognitive function in GM using mini-mental state examination (MMSE) and Montreal cognitive assessment (MoCA) in AD patients. The result provides a high susceptibility value in the PUT and CN and a low susceptibility value in the RN region for AD patients. Additionally, the paper suggested that the susceptibility value of the left caudate nucleus could be employed as a marker to diagnose disease severity in mild and moderate AD. Moreover, the paper indicated that right and left brains do not always provide an iron alteration in a parallel mode.

Li et al. [51] have studied changes based on cerebral blood flow (CBF) and iron deposition in the brain. They observed Gd had the highest QSM values for the HC group. In contrast, HP had the lowest susceptibility value. This concluded that the distribution of iron varies with different brain regions. Also, the PUT brain region showed the highest QSM values using reduced blood perfusion, indicating that iron deposition during AD progression differs from iron deposition in healthy elderly brains. Hence, PUT in AD patients might be sensitive to iron deposition. QSM values in the PUT can be used as an imaging biomarker for early AD diagnosis. Similarly, Au et al. [52] have discussed the QSM technique to measure the microstructural composition of the hippocampal fimbria. The paper suggested that WM played an equally significant role in AD pathogenesis as GM and helped to examine the possibility of hippocampus fimbria as a tool for AD diagnosis. Tissue voxels have paramagnetic and diamagnetic properties, which showed different susceptibility values. Moreover, high susceptibility can significantly influence the overall QSM value. The lipids and proteins have a diamagnetic property and can measure the magnetic susceptibility of WM. As a result, normal hippocampus fimbriae delivered negative QSM values as white matter in the control group. In contrast, an increasing QSM positive value in the AD case was found, implying susceptibility value from the early to late stages. However, a larger dataset will be required to confirm these findings as the article has taken a small number of subjects.

In [53], the author has estimated the cerebral venous oxygen level in AD. Blood oxygen levels in the cerebral veins of AD patients were initially measured using QSM. This study observed that the bilateral cerebral veins' oxygen levels in the AD group were generally lower than those in the confirmed AD groups as the amount of deoxyhemoglobin increased. However, these significant differences were observed only in the left dentate nucleus and bilateral thalamus region. The paper has used the diameter of the blood vessel as an additional important factor in diagnosing AD patients. Moreover, finding marked that small blood vessels are more susceptible to the partial volume and slice effects during the measuring process and cause variations in the measurement results. Furthermore, Ryota Sato et al. [54] proposed a new diagnostic index for AD based on a hybrid sequence of QSM and VBM. The hybrid sequence was extracted using 3T MRI and a 3D multi-echo GRE sequence and obtained T1-weighted images for VBM and the phase images for QSM, respectively. Afterward, a linear support vector machine was used to generate the index from particular voxels in QSM and VBM images. Further, the proposed QSM and VBM-based index can be used as a biomarker to track the progression of AD. The result yielded better diagnostic performance between AD and HC as well as MCI and HC. In [55], author have discussed early-onset Alzheimer's disease (EOAD) using QSM. The study considered iron allocation with EOAD subtypes in cognitive HC (those at risk of developing AD). Seven ROIs within limbic and deep grey nuclei structures were selected for the QSM voxelwise analysis. The result delivered the highest QSM values in HpSpMRI (i.e., EOAD subtypes) and LPMRI brain regions of deep gray nuclei and limbic structures, respectively.

*4.2 Parkinson's Disease*

Dopamine is a message carrier that flows in the brain and nervous system responsible for controlling and coordinating body movement. Loss of nerve cells, mainly in SN, reduces dopamine quantity, which causes Parkinson's Disease (PD) [56, 57]. Neuronal death is believed to be caused due to oxidative stress. In contrast, highly reactive hydroxyl radical, which is also formed due to the free iron particle, is supposed to be the byproduct of the oxidative deamination of dopamine [58]. Abnormal iron level in BG related to PD efficiently quantified by the QSM. In article [59], Shahmaei et al. estimated the 15 HC and 30 PD patients and observed iron deposition in the BG region. Chen et al. [60] also investigated the QSM technique in deep gray nuclei and SN regions. The result shows the iron deposition gradient in GP-fascicule nigral (FN)-SN pathways for 33 PD and 26 HC subjects. Other previous studies are mentioned in Table 2.

Table 2 Application of QSM in PD

| Citation | Method | Database | `Resulted higher Iron accumulation regions |
|---|---|---|---|
| [61] | HPF [62] and TKD | PD-09 HC-11 | PC |
| [63] | PDF and MEDI | PD-21 HC-21 | SN |
| [64] | TSVD-SHARP and TKD | PD-20 HC-30 | SNc |
| [65] | MEDI | PD-47 HC-47 | SNc |
| [66] | Laplacian unwrapping and V-SHARP | PD-62 HC-40 | DN and RN |
| [67] | SHARP and Iterative LSQR [68] | PD-87 HC-77 | SN |
| [69] | MEDI and Laplacian boundary value | PD-20 HC-20 | SN and RN |
| [70] | V-SHARP STAR-QSM | PD-32 HC-30 | SN, RN and CN |
| [71] | V-SHARP and Iterative LSQR | PD-39 HC-28 | SN |
| [72] | - | PD-104 HC-45 | SN |

Lotfipour et al. [61] have applied high-resolution magnetic susceptibility mapping and found the distribution of dopaminergic neurons homogeneous inside the SN. In this paper, QSM detected the susceptibility differences of spatial variations within the SN region for PD patients. Moreover, the iron content of the methodology provides a marker to measure the high susceptibility value in the SN region. However, a small dataset was employed; therefore, the outcome can give a different result. Similarly, Murakami et al. [63] have considerably higher QSM and R2* values in the SN because iron with paramagnetic properties influences MR imaging's susceptibility and relaxation contrasts. Also, the author observed no difference in other brain regions between patients and HC. Paper small datasets and difficulty getting a histopathologic confirmation of PD patients could affect the validity of result analyses. In [64], the paper has discussed R2, R2*, and QSM map to compute the sensitivity and specificity that differentiate the HC and PD individuals. The final results confirmed that the substantia nigra pars compacta (SNc) structure displayed a high iron level in PD subjects. Similarly, Du et al. [65] have explored the iron content to assess the significant susceptibility value changes in both the right and left midbrain (including SNc). They concluded that in the QSM, the higher dynamic value of susceptibility makes it a reliable biomarker compared to R2* for disease severity detection. However, QSM may be influenced by other parameters like calcium, lipid, or myelin content.

Guan et al. [66] have identified susceptibility alteration and compared the underlying WM differences of QSM and DTI. Specifically, these alterations were observed in brain areas like the frontal, parietal, and temporal lobes. Additionally, the right cerebellar hemisphere and the left temporal lobe displayed relative WM changes on QSM and DTI. However, this study estimated mean magnetic susceptibility rather than susceptibility anisotropy. Cheng et al. [67] have selected 40 radiomic features using a feature ensemble algorithm and noted that these SN region features could help diagnose the PD patient. Moreover, the author used the support vector machine (SVM) model to classify PD patients and HC subjects which yielded an accuracy of 88 percent. Basukala et al. [69] have proposed a segmentation method for QSM images to segment SN and RN regions. They compared algorithm-derived segmentation (based on level set and watershed transform) and expert manually-derived segmentation approaches. These approaches demonstrated that the proposed algorithm performs better than expert manual-based segmentation. In addition, they observed high susceptibility values in the SN region for PD subjects.

Xu et al. [70] have analyzed peripheral inflammatory cytokines, iron metabolism, and brain iron deposition in PD subjects using QSM. They compared the PD group to the HC and the advanced-stage PD group to the early-stage PD group. Both groups exhibited significantly higher total QSM values for bilateral ROIs. In addition, results illustrated that total QSM values in PD patients had a significant inverse connection with serum IL-1b concentration for bilateral ROIs. However, no direct link was observed between serum IL-1b concentrations and serum ferritin, iron, etc., in PD patients. Wang et al. [71] compared the QSM and positron emission tomography (PET) techniques for early diagnosis of PD. They investigated that imaging of swallow tail sign (STS) on MRI for nigrosome-1 that may play a similar role for PET in identifying dopaminergic degeneration. Kang et al. [72] investigated QSM-based magnetic susceptibility value (MSVs) and radiomics characteristics. The paper used multivariable logistic regression (MLR) and SVM models to diagnose PD. In addition, the correlation between MSVs, radiomics characteristics, and Montreal Cognitive Assessment scores (MoCA) was examined. The highest area under the curves (AUCs) was found in the MSVs case on the right side of SNc, whereas at the right SN region for radiomics characteristics.

*4.3 Multiple Sclerosis*

Haemoglobin is a significant protein in erythrocytes, and its primary source can be iron. Many researchers suggest that iron overload or its deficiency in the long-term causes neural cell death in some parts of the brain and is responsible for demyelination which slows down the message sent capability of the axon. Although the exact cause of multiple sclerosis (MS) is not completely clear yet, it believes that myelin (membranes that cover an axon) destruction is responsible for this. Several studies [73] have suggested that the proper iron level is essential for myelin production. Its unbalancing is assumed to be the reason for MS. Similarly, Winkler et al. [74] have examined solid and shell susceptibility using QSM and phase imaging techniques. The result displayed that QSM finds to be accurate in investigating MS, whereas phase imaging fails to do that.

Table 3 Application of QSM in MS

| Citation | Method | Database | Resulted higher Iron accumulation regions |
|---|---|---|---|
| [75] | MEDI | MS-68 HC-23 | BG and GP |
| [76] | - | MS-32 | NAWM |
| [77] | Laplacian Unwrapping, SHARP and LSQR | MS-24 | Periventricular areas |
| [78] | MEDI | MS-29 | Enhanced-with-Gd lesions |
| [79] | SHARP and HEIDI | MS-600 HC-250 | BG and GP |
| [80] | MEDI | MS-30 | Rim+ lesions |

Langkammer et al. [75] have employed a three-dimensional GRE sequence to generate the susceptibility and R2* maps. The paper evaluated mean susceptibilities and R2* maps using variance analysis in the BG region. Further, the result showed that QSM is more sensitive than R2* maps and yielded higher magnetic susceptibility in MS individuals. Also, Chen et al. [76] have selected 32 clinically confirmed MS patients who had experienced two MRI exams. The paper utilized hyperintensity on T2-weighted images to determine these lesions. Moreover, in MS lesions' magnetic susceptibility rapidly increased from enhanced to non-enhanced and achieved a higher susceptibility value than NAWM. Result analysis of lesion obtained from t-test with intracluster correlation adjustment and Bonferroni correction. In contrast, Li et al. [77] have selected 306 lesions of 24 MS patients. QSM and R2* mapping techniques were employed to diagnose MS. Author calculated each lesion using manually created lesion masks, R2*, and susceptibility values. Further, they compared lesion groups to demonstrate various contrast patterns. Also, classify them depending on the image intensity or the anatomical locations.

Zhang et al. [78] studied longitudinal changes in quantitative susceptibility values of new enhanced-with-Gd lesions. Paper estimated these changes in MS lesions in 29 patients (have a follow-up MRI within two years) using generalized estimating equations (GEE) model and t-test. Results found an increment in lesion growth from the enhanced to the not-enhanced ROI region. Similarly, an article by R.Zivadinov et al. [79] suggested that iron deposition is associated with MS in the deep GM region. The susceptibility of GM was examined by ROI and voxel-based approaches such as QSM and MRI, which later assist in evaluating clinical outcomes. The paper demonstrated that subjects with MS exhibited lower susceptibility in the Th region and higher susceptibility in BG compared to HC. Similarly, Kaunzner et al. [80] have detected a higher susceptibility value in rim+ lesion than rim- the lesion region using QSM. Therefore, paper suggested that region could be used as biomarkers to detect MS patients' risk levels.

*4.4 Aging*

In the aging brain, iron deposition is not normal, which might be the reason for memory impairment in old age. Besides that, myelin and myelin sheath contain WM protein and exhibits iron accumulation which causes aging. A technique like QSM is used to measure these biomental depositions.

Table 4 Application of QSM in Aging

| Citation | Method | Database | Resulted higher Iron accumulation regions |
|---|---|---|---|
| [81] | L1 and L2 regularization | YA-11 EA-12 | PUT and GP |
| [82] | PDF and L1-QSM | YA-11 EA-12 | GP and PUT |
| [83] | SHARP and MEDI | YA-20 EA-20 | PUT, Caudate, STN SN, RN, HP and DN |
| [84] | - | YA-30 EA-10 | STr, RN, SN |
| [85] | SHARP and HEIDI | HC- 213 | CN, PUT, GP, Th, RN, SN, and DN |
| [86] | SHARP and TKD | HC-623 | GP, Th and pulvinar Th |

*\* Sub-thalamic nucleus*

Bilgic et al. [81] have employed L1 and L2 regularization algorithms of QSM that were used for dipole fitting and convex optimization [88] to eliminate the background phase and get the brain tissue's susceptibility value, respectively. Iron accumulation in the aging brain of the adult (younger adults (YA), elderly adult (EA)) was measured, and results indicate higher iron in striatal and brain stem ROIs of the elderly than in young adults. Poynton et al. [82] investigated quantifying susceptibility by inversion of a perturbation model, this model use kernel in the spatial domain that associates the phase to susceptibility. The model compares to QSM algorithms and yields a more accurate magnetic susceptibility. Similarly, a comparative study of QSM and R2* has been performed by Betts et al. [83] that detects brain non-haem iron accumulation, calcification, demyelination, a vascular lesion. The QSM and R2* algorithm yielded effective age-related differences in the superior frontal region, HP, and GM for the aging brain. However, the following techniques provided distinguishing results in GP, Th, occipital, and cortex brain regions. Anatomical alterations in aging brains have been observed by Keuken et al. [84]. MRI can be used to measure these variations. With the advancement of MRI, the field strength [87] above 3T or some 7T can be exploited, which helps to take even the picture of small subcortical parts. Moreover, the paper has explored different brain regions of aging subjects using volumetric, spatial, and quantitative MRI parameters. The findings demonstrate that aging showed performance varied with different brain areas.

Zhou et al. [85] aimed to identify the relationship between iron accumulation and the glymphatic system's role in the normal aging brain. The study revealed that the local brain region iron quantity was connected with the glymphatic system. Li et al. [86] evaluated how image resolution affects QSM quantification using the simulation model. Data from 623 HC subjects ranging from 20 to 90 years old was taken. Further, QSM data were semi-automatically segmented using a full-width half-maximum threshold analysis. Moreover, correlation with the age of each subcortical gray matter nuclei was assessed by susceptibility, total iron content, etc., for the global and RII (high iron content regions) analysis. In the case of global analysis, mean susceptibilities of CN, PUT, RN, SN, and DN were associated positively, whereas the Th region and other volume regions correlated negatively with age. Besides GP, Th, and pulvinar thalamus, all the other structures delivered increment of total iron content despite volume declines with age.

*4.6 Discussion*

MRI, a non-invasive technique, is practiced mainly for soft tissue contrast and structural changes. One reason for structural and anatomical alteration in the brain is iron accumulation which is efficiently measured using the difference in magnetic susceptibility value. The multi-GRE sequence is suitable for collecting these variations, and the techniques mostly preferred for this are SWI, QSM, etc. Both methods work on susceptibility mapping; however, SWI only provides contrast in the image, whereas QSM quantifies the susceptibility difference of neural contrast.

Table 5 Application of QSM for predicting the biomarker of neural disorder

| Neural Disorder | ROIs | | | | | | | | | | |
|---|---|---|---|---|---|---|---|---|---|---|---|
| AD's patients | GM | WM | CSF | GP | PUT | CN | Th | SN | DN | RN | HP |
| [46] | ✓ | ✓ | - | ✓ | ✓ | ✓ | ✓ | - | - | - | - |
| [47] | - | - | - | ✓ | ✓ | ✓ | ✓ | - | - | - | - |
| [48] | ✓ | ✓ | - | - | - | - | - | - | - | - | - |
| [49] | ✓ | ✓ | ✓ | - | - | - | - | - | - | - | - |
| [50] | - | - | - | ✓ | ✓ | ✓ | ✓ | ✓ | ✓ | ✓ | - |
| [51] | - | - | - | ✓ | ✓ | - | - | - | - | - | - |
| [52] | - | - | - | - | - | - | - | - | - | - | ✓ |
| [53] | - | - | - | - | - | - | - | ✓ | - | ✓ | - | - |
| [54] | ✓ | ✓ | - | - | - | - | - | - | - | - | - |
| [55] | - | - | - | ✓ | ✓ | ✓ | ✓ | - | - | - | ✓ |
| PD's patients | GM | WM | PC | GP | PUT | CN | Th | SN | SNc | RN | |
| [61] | - | - | ✓ | - | - | - | - | ✓ | - | ✓ | |
| [63] | - | - | - | - | ✓ | ✓ | ✓ | ✓ | - | ✓ | |
| [64] | ✓ | ✓ | - | ✓ | ✓ | ✓ | ✓ | ✓ | ✓ | ✓ | |
| [65] | - | - | - | - | - | - | - | - | ✓ | - | |
| [66] | - | - | - | ✓ | ✓ | - | ✓ | ✓ | ✓ | ✓ | |
| [67] | - | - | - | - | - | - | - | ✓ | - | - | |
| [69] | - | - | - | - | - | - | - | ✓ | - | - | |
| [70] | - | - | - | - | - | ✓ | - | ✓ | - | ✓ | |
| [71] | - | - | - | - | - | - | - | ✓ | - | - | |
| [72] | - | - | - | - | ✓ | ✓ | - | ✓ | - | - | |
| MS patients | GM | | WM | | | | | Periventricular | | | |
| [75] | ✓ | | ✓ | | | | | - | | | |
| [76] | - | | ✓ | | | | | - | | | |
| [77] | - | | - | | | | | ✓ | | | |
| [78] | - | | ✓ | | | | | - | | | |
| [79] | ✓ | | ✓ | | | | | - | | | |

| | [80] | - | ✓ | | | | - | | | | |
|---|---|---|---|---|---|---|---|---|---|---|---|
| *Aging* | | **HP** | **WM** | **BG** | **GP** | **PUT** | **CN** | **Th** | **SN** | **DN** | **RN** |
| | [81] | - | ✓ | ✓ | ✓ | ✓ | - | ✓ | ✓ | - | - |
| | [82] | - | ✓ | - | ✓ | ✓ | - | ✓ | - | - | - |
| | [83] | ✓ | - | - | - | ✓ | - | ✓ | ✓ | ✓ | ✓ |
| | [84] | - | - | - | - | - | - | - | ✓ | - | ✓ |
| | [85] | - | - | - | ✓ | ✓ | ✓ | ✓ | ✓ | ✓ | ✓ |
| | [86] | - | - | - | ✓ | ✓ | ✓ | ✓ | ✓ | ✓ | ✓ |

Many neurodegenerative diseases, such as AD, PD, MS, aging effect, etc., can be used to differentiate between HC and patients suffering from neurological diseases by QSM technique. As stated in Table 5, AD is, specified in GM, WM, GP, PUT, CN, and Th, exhibits high iron accumulation in some past studies, whereas few papers also included the CSF, SN, DN, and RN brain region.

Previous work on PD displayed the highest checkmark in the SN region, whereas the RN brain part was also indicated as an iron deposition region in the past four studies. PUT, Th, and SNc were selected and showed high iron quantities in past papers. In the same way, some papers confirmed displays of GM, WM, and PC areas that represented high susceptibility difference regions compared to the normal controls.

Similarly, in the case of MS lesions that extend from enhanced to non-enhanced regions, mostly in WM, five studies indicated this part. Some previous works also mentioned the GM region and periventricular brain section.

High iron quantity is assumed to be linked with aging in distinct brain regions, which alter the human brain regions and show high susceptibility difference values in the PUT, Th, and SN brain parts. GP and RN are also shown by some papers which contain iron accumulation. A checkmark intimated WM and RN brain regions in two studies.

*5. Future Scope*

QSM techniques include MRI images, but these image details are sensitive to head motion. The article also assumed isotropic magnetic susceptibility identified as a restriction in the case of anisotropic susceptibility. Water-fat separation during imaging is another dilemma confronted by the researcher, solved by applying a water/fat imaging sequence, which will give water-only and fat-only images.

The first stage of the QSM pipeline is phase unwrapping. The laplacian method is frequently used for this unwrapping, which holds complex equations. For this reason, equations convert into the Fourier domain to have the accessible environment that avails in solving the equations, usually done by fast Fourier transform (FFT). However, this affects the boundary condition, which can be resolved using the second time FFT or a different one but this will take more computation time. Background field removal is another primary step in concocting susceptibility images. SHARP and RESHARP have limitations as some local fields vanish at the boundary and offer surplus error in the high susceptibility region. PDF can avoid these constraints, making susceptibility sources fit outside the generated ROI and providing background details. However, some aliasing effects are detected. Hence, these limitations address the possibility of extending the study to a deep learning area, which helps in diminishing the residual field errors.

For the susceptibility inversion, COSMOS is assumed as a gold-standard technique for this, but it requires multiple orientations that make its practicability difficult. A single orientation method, such as TKD and MEDI, can be employed for the inversion problem. TKD and TSVD define the threshold value, which needs to be adjusted to move this methodology in the demerit list. MEDI utilizes L1 norm minimization that increases the complexity. Similarly, TVSB demands extra computational time. Therefore, better, fast, and practically easy algorithms require to improve the ill-posed inversion problem. The deep learning model is convenient in resolving the ill-posed field-to-source inversion problem; future practices might be added to this reference.

In addition, the following study is bound to iron quantification; therefore, other elements, calcium, magnesium, phosphorus, etc., provide new opportunities for thoughtful work in this region. Also, the paper application mainly focuses on cognitive decline; so additional work can be explored for brain tumors, brain hemorrhages, etc. QSM application can also be explored for inhomogeneity detection in other parts of the body like: kidney, liver, lungs, etc.,

*6. Conclusion*

This review paper introduced a systematic review of the literature on QSM processing steps and its applications in diagnosing neural disorders. QSM images processed raw phase images which are unwrapped by the Laplacian phase and path-following phase unwrapping methods. These methods rectify the phase jumps or phase discontinuity. After that, background field removal is employed to eliminate the unwanted image part. Background field removal is performed using SHARP, RESHARP, and the PDF algorithm. Further, the susceptibility inversion step assists in reconstructing the QSM image. This paper utilizes COSMOS, MEDI, TVSB, TVSB, and TKD methodology to perform inversion operations.

This review also presents an in-depth review of the applications of QSM images in several neural disorders to find the region of interest of iron accumulation. This review will effectively help the researchers and clinicians to find the applications of QSM in early diagnosing of cognitive decline and give the future directions for the further improvement in this field.

**Compliance with Ethical Standards**

**Funding:** This study does not contain any funding source.

**Ethical approval**: This article does not contain any studies with human participants performed by any of the authors.

**Informed Consent:** As this article does not contain any studies with human participants or animals performed by any of the authors, the informed consent is not applicable.

**Conflict of Interest**: The authors declare no competing interests.

**Data Availability Statement:** No datasets were generated or analyzed during the current study.